\documentclass[
reprint,
twocolumn,
superscriptaddress,
amsmath,amssymb,
aps,
prx,
floatfix,
]{revtex4-1}

\usepackage{subcaption}
\usepackage{graphicx}
\usepackage{graphics}
\usepackage{amsmath}
\usepackage{amssymb}
\usepackage{amsfonts}
\usepackage{amsfonts}
\usepackage{array}
\usepackage{array}
\usepackage{colortbl}
\usepackage{dsfont}
\usepackage{braket}
\usepackage{color}
\usepackage{booktabs}
\usepackage{braket,slashed}
\usepackage[mathscr]{euscript}
\definecolor{darkblue}{rgb}{0, 0, 0.8}
\usepackage[colorlinks=true, breaklinks=true, linkcolor=red, citecolor=blue, urlcolor=blue]{hyperref} 
\usepackage{hyperref}
\usepackage{natbib}
\usepackage{xfrac}
\usepackage{bm}
\usepackage{kantlipsum}
\usepackage{enumitem}
\usepackage{tikz}
\usepackage{framed}
\usepackage{graphicx}
\usepackage{cleveref}
\usepackage{adjustbox}
\usepackage{xcolor}

\allowdisplaybreaks[1]

\newcommand{\code}[1]{\texttt{#1}}

\newcommand{\be}{\begin{equation}}
\newcommand{\ee}{\end{equation}}

\newcommand{\bea}{\begin{eqnarray}}
\newcommand{\eea}{\end{eqnarray}}

\usepackage{float}
\usepackage{multibib,ulem}
\newcites{app}{Appendix References}
\usepackage{hyperref}

\begin{document}

\title{Topological Criticality with Charge-\(3e\) Fluctuations
in the \(\mathrm{SU}(3)\) Triangular-Lattice
Hofstadter--Hubbard Model}

\author{Yang Liu}
\affiliation{Department of Physics, The Hong Kong University
of Science and Technology, Clear Water Bay, Kowloon,
Hong Kong, China.}

\author{Ya-Hui Zhang}
\affiliation{Department of Physics and Astronomy,
The Johns Hopkins University, Baltimore, MD, USA}

\author{Xue-Yang Song}
\affiliation{Department of Physics, The Hong Kong University
of Science and Technology, Clear Water Bay, Kowloon,
Hong Kong, China.}

\begin{abstract}
A transition between an integer quantum Hall insulator and a chiral spin liquid can become critical without closing the single-electron gap, raising the question of which composite excitations govern the change in topological order. We investigate this question in the \(\mathrm{SU}(3)\) triangular-lattice Hofstadter--Hubbard model at one particle per site and magnetic flux \(\pi/3\) per triangle. Using infinite density matrix renormalization group calculations, we identify a weak-coupling integer quantum Hall phase 
and an interaction-driven 
\(\mathrm{SU}(3)_1\) chiral spin liquid.
Near the transition, both the neutral and charge-\(3e\)
correlation lengths increase substantially, whereas the
single-electron correlation length remains short,
providing evidence for a Chern--Simons--Higgs critical theory
, where a gauge-invariant monopole carries the quantum numbers of a fermionic \(\mathrm{SU}(3)\)-singlet trion. 
We discuss the implications for doping the system and recent experiments on doping $\nu=-2/3$ fractional quantum anomalous Hall in tMoTe$_2$.
\end{abstract}

\maketitle

\section{Introduction}
\label{sec:intro}

Quantum phase transitions between topologically distinct states
need not be accompanied by conventional symmetry breaking.
Instead, the low-energy theory can involve fractionalized excitations coupled to emergent gauge fields, whose dynamics reorganize the topological order and quantized responses of the neighboring phases
\cite{WenWu1993,Lee2018FCI}.
Chiral spin liquids (CSLs) offer a particularly useful setting for exploring such transitions: they combine a gapped bulk with
anyonic excitations and chiral edge modes
\cite{Kalmeyer1987,WenWilczekZee1989}. A central question is which physical excitations become critical when a CSL is reached from an electronic insulator. 
Particularly intriguing are transitions at which microscopic electrons remain gapped while collective
or composite excitations become gapless.

The transition between an integer quantum Hall (IQH) insulator
and a chiral spin liquid (CSL) provides a concrete setting for
addressing this problem. Recent infinite density matrix
renormalization group (iDMRG) calculations of the
\(\mathrm{SU}(2)\) triangular-lattice Hofstadter--Hubbard model
found evidence for a continuous transition with prominent
spin-singlet, charge-neutral current fluctuations, while spin
and single-electron correlations remain short-ranged
\cite{Divic2026critical}. 
Complementary theoretical and numerical work connects this topological transition to
low-energy charged composites and superconducting tendencies upon doping\cite{divic_anyon,chen_2026topological,kuhlenkamp_2025robust}, realizing the celebrated anyon superconductivity\cite{Laughlin1988a,Laughlin1988b,Song2021,Song2023,zhang2025charge}. The enhancement of charge-\(2e\) correlations, without a comparable enhancement of single-electron correlations, provides further evidence for composite charge dynamics.Together, these results motivate a sector-resolved investigation of topological criticality beyond the two-flavor setting.

The \(\mathrm{SU}(N)\) Hofstadter--Hubbard family provides a natural extension. Earlier slave-rotor studies showed that background magnetic flux can stabilize CSLs in multiflavor Hubbard models \cite{Chen2016Synthetic, Hermele2009},
and the IQH--CSL framework admits a corresponding
\(\mathrm{SU}(N)\) generalization \cite{divic_anyon,zhang_2025phases}.
The local flavor-singlet composite contains \(N\)
electrons and therefore has bosonic statistics for even
\(N\) and fermionic statistics for odd \(N\), consistent
with the electronic spin/charge relation
\cite{Seiberg2016Gapped}.

In this work, we study the corresponding transition in an
\(\mathrm{SU}(3)\) system. Three flavors introduce a qualitative
distinction: the local flavor-singlet composite
\[
\mathcal{T}_i=c_{i,1}c_{i,2}c_{i,3}
\]
carries charge of magnitude \(3e\) and is fermionic.
It is therefore possible for an odd-charge composite sector to become critical even when the single-electron sector remains gapped. In a parton description, this possibility is tied to the anyon structure of the CSL. Integrating out gapped flavor-carrying partons produces a level-three Chern--Simons term for an emergent gauge field.
Gauge invariance then requires a unit monopole to be dressed by three charged matter fields\cite{kapustin}. The resulting operator has the charge and flavor quantum numbers of a singlet trion.
This identifies a concrete physical channel through which fractionalized degrees of freedom can manifest themselves at the transition.

We investigate the \(\mathrm{SU}(3)\) triangular-lattice
Hofstadter--Hubbard model at one particle per site and flux \(\pi/3\) per triangle. Charge pumping and
\(\mathrm{SU}(3)\)-resolved entanglement spectra identify the neighboring IQH and candidate \(\mathrm{SU}(3)_1\) CSL phases. We then resolve the matrix product state transfer spectrum by charge to determine which sectors develop
long correlation lengths. Near the transition, the neutral and charge-\(3e\) correlation lengths grow, while the charge-\(e\) correlation length remains short. Their dependence on interaction strength, bond dimension, and cylinder circumference supports a continuous-transition scenario with low-energy trion fluctuations.

Our study thus complements the \(\mathrm{SU}(N)\) investigation by focusing on a fermionic composite charge sector.
We distinguish the numerical evidence for enhanced
charge-\(3e\) correlations from its proposed monopole
interpretation, and discuss the additional extrapolations needed to establish the two-dimensional critical behavior.
Finally, we outline possible implications for doped
\(\mathrm{SU}(3)\) chiral spin liquids.

\section{Topological transition and candidate critical theory}
\label{sec:topo_transition}

\begin{figure*}[htp]
    \centering
    \includegraphics[width=0.9\textwidth]{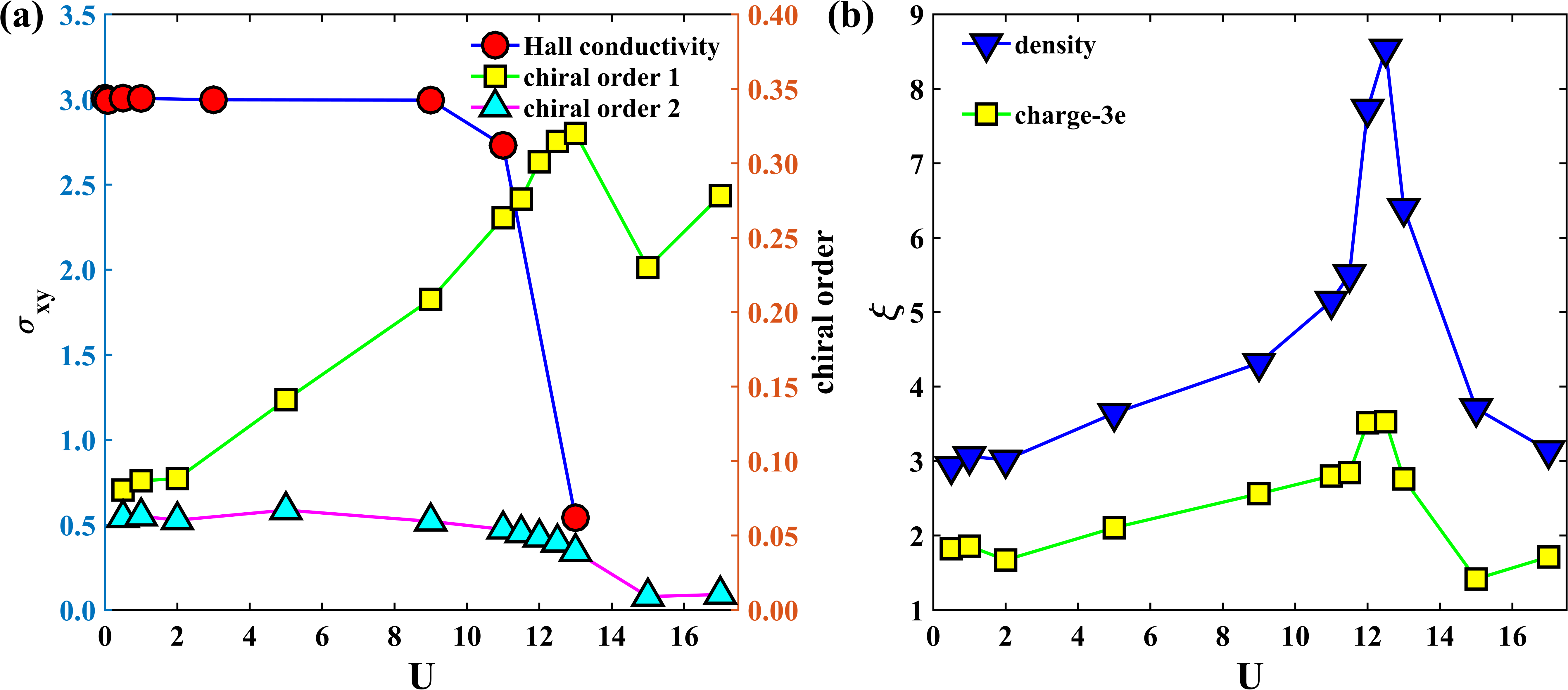}
    \caption{
    Diagnostics of the interaction-driven transition in the
    \(\mathrm{SU}(3)\) Hofstadter--Hubbard model.
    At weak coupling, charge pumping is consistent with an IQH
    phase of Hall conductance \(3e^2/h\).
    Near the transition, the pumped-charge response departs
    from its weak-coupling value.
    The chirality contributions resolved by local occupation
    sectors evolve as increasing repulsion favors single
    occupancy. These measurements are considered together
    with the entanglement spectrum and correlation lengths
    to identify the neighboring phases and transition region.
    }
    \label{fig:phase_transition}
\end{figure*}

We consider an interaction-driven transition between an IQH
state with total Chern number \(C=3\) and an
\(\mathrm{SU}(3)_1\) CSL. Research on the triangular-lattice Hofstadter--Hubbard model based on parton mean-field theory already exists
\cite{zhang_2025phases}.In the noninteracting limit,
each of the three flavors fills the lowest Hofstadter band,
giving a Hall conductance of \(3e^2/h\). Increasing the
onsite repulsion suppresses charge fluctuations and can
stabilize a CSL with neutral chiral edge excitations.

To describe this transition, we use the parton decomposition
\[
c_\alpha=b f_\alpha,\qquad \alpha=1,2,3,
\]
where the chargon \(b\) carries the physical charge and the fermionic
parton \(f_\alpha\) carries the flavor quantum number.
The decomposition introduces an emergent \(\mathrm{U}(1)\)
gauge redundancy, under which the two partons carry
opposite gauge charges. We consider a mean-field state
in which each spinon flavor occupies a Chern band with
Chern number \(+1\).

The resulting \(\mathrm{SU}(3)_1\) CSL has three anyon
types,
\[
\mathbf{1},\qquad \mathbf{3},\qquad \bar{\mathbf{3}},
\]
with topological-spin phases \(1\), \(e^{2\pi i/3}\),
and \(e^{2\pi i/3}\), respectively
\cite{DiFrancesco1997Conformal,Moore1989Naturality,
Witten1989Quantum}.
Here the representation labels describe the flavor
quantum numbers of the anyons, while their fusion rules
form a \(\mathbb{Z}_3\) structure.

Integrating out the gapped spinons generates a level-three
Chern--Simons term. A candidate continuum description
of the charge-sector transition is
\begin{equation}
\label{eq: critical}
\begin{split}
\mathcal{L}_{\mathrm{critical}}
={}&
|\mathcal{D}_{a-A}\psi|^2
-r|\psi|^2-u|\psi|^4
\\
&+\frac{3}{4\pi}
\epsilon^{\mu\nu\rho}a_\mu\partial_\nu a_\rho
+\cdots ,
\end{split}
\end{equation}
where
\(\mathcal{D}_{a-A,\mu}=\partial_\mu-i(a_\mu-A_\mu)\),
\(a_\mu\) is the emergent gauge field, \(A_\mu\) is the
electromagnetic probe field, and \(\psi\) describes
the low-energy chargon mode. 

When the chargon is gapped, the spinon construction
describes the CSL. In the Higgs phase, the relative
gauge field \(a-A\) is locked, recovering the IQH charge
response. Thus the transition concerns the reorganization
of the fractionalized charge sector. 


A central consequence of the Chern--Simons term is the
gauge charge carried by a monopole insertion.
A bare unit monopole \(\mathcal{M}_0\) must be dressed by three
charged matter fields to form a gauge-invariant operator
\cite{kapustin}:
\[
\mathcal{M}=\mathcal{M}_0\psi^3.
\]
This operator carries physical charge of magnitude \(3e\)
and has fermionic statistics. Its charge and flavor
quantum numbers agree with those of the local singlet
trion \(\mathcal{T}=c_1c_2c_3\), suggesting the critical trion fluctuations. Note that there is no low-energy operator in the effective theory eq \eqref{eq: critical} with quantum numbers identical to those of a single electron. This indicates the single electron gap remains open across the transition\cite{divic_anyon}.

Apart from the preceding $SU(2)$ study\cite{divic_anyon},the broader framework of deconfined criticality
\cite{Senthil2004}, boson--fermion dualities
\cite{Seiberg2016}, and gauge-field descriptions of
superconductor--insulator and anyonic transitions
\cite{Fisher1990,Pryadko1998} motivates examining such
a Chern--Simons--matter theory. We hence investigate whether the microscopic \(N=3\) transition is continuous and 
its universality class. Next we report numerical findings
to identify the neighboring phases and test whether
the neutral and trion sectors become long-ranged
while single-electron correlations remain short,
as suggested by the parton picture
\cite{divic_anyon,zhang_2025phases}.

\section{Model and numerical results}
\label{sec:model_results}

We study the trangular lattice Hamiltonian
\begin{equation}
\label{eq:hamiltonian}
\begin{split}
\mathcal{H}
={}&-t\sum_{\langle ij\rangle}\sum_{\alpha=1}^{3}
\left(
e^{-i\theta_{ij}}c^\dagger_{i,\alpha}c_{j,\alpha}
+\mathrm{H.c.}
\right)
\\
&+U\sum_i\sum_{\alpha\ne\beta}
n_{i,\alpha}n_{i,\beta}.
\end{split}
\end{equation}
The Peierls phases produce flux \(\pi/3\) through each
elementary triangle. We impose the average density
\[
\frac{1}{N_s}\sum_i\sum_{\alpha=1}^{3}
\langle n_{i,\alpha}\rangle=1,
\]
where \(N_s\) is the number of sites. This density
corresponds to one-third filling of the three-flavor
local Hilbert space.


At \(U=0\), each flavor fills the lowest Hofstadter band,
and the ground state is an IQH insulator with
\(\sigma_{xy}=3e^2/h\). At stronger coupling, charge
fluctuations are suppressed. Whether the resulting
state is a CSL must be established from its many-body
properties. 
alone.

We set \(t=1\) and use iDMRG to compute ground states
on infinite cylinders. The analysis combines charge
pumping, occupation-resolved chirality, entanglement
spectra, and transfer-matrix correlation lengths.
Geometry conventions and supplementary results are
given in Appendices~\ref{sec:lattice}
and~\ref{sec:pump_charge_correlation_different_Ly}.


\subsection{Phase diagnostics}

We first examine the charge response under flux insertion.
At weak coupling, the pumped charge is consistent with
\(\sigma_{xy}=3e^2/h\), as shown in
Fig.~\ref{fig:phase_transition}(a). Around \(U\simeq12\),
the response departs from its weak-coupling value,
locating a region in which the IQH state changes.
Details of the pumping procedure are provided in
Appendix~\ref{sec:pump_charge_correlation_different_Ly}.

We also measure the generalized three-site chirality
\[
\chi_{123}=i(P_{123}-P_{132}),
\]
where \(P_{132}=P_{123}^{\dagger}\). In the singly
occupied subspace, \(P_{123}\) cyclically permutes the
three local flavors 
(details in Appendix~\ref{sec:su3_formalism}).

The local one- and two-particle sectors transform as
\(\mathbf{3}\) and \(\bar{\mathbf{3}}\), respectively.
We focus on chirality contributions associated with
\(\mathbf{3}\otimes\mathbf{3}\otimes\mathbf{3}\) and
\(\mathbf{3}\otimes\mathbf{3}\otimes\bar{\mathbf{3}}\),
denoted in the figures as chiral order 1 and chiral
order 2. These labels specify local occupation sectors,
not irreducible decompositions of the three-site
Hilbert space.

The contribution from three singly occupied sites
increases as repulsion favors single occupancy, whereas
contributions involving doubly occupied sites are
suppressed. Their evolution is consistent with the
development of a Mott regime. Because the applied flux
already breaks time-reversal symmetry, a nonzero
chirality is not by itself evidence for a CSL.
We therefore combine this diagnostic with the
entanglement spectrum discussed below.

The central question is which physical sectors develop
long-ranged correlations near the transition.
We extract correlation lengths from the matrix product
state transfer matrix. If \(\lambda_0\) is its dominant
eigenvalue and \(\lambda_q\) is the leading relevant
eigenvalue in a sector \(q\), then
\begin{equation}
\label{eq:transfer_xi}
\xi_q^{-1}
=-\frac{1}{\ell}
\ln\left|\frac{\lambda_q}{\lambda_0}\right|,
\end{equation}
where \(\ell\) is the length advanced by one application
of the transfer matrix. In the neutral sector,
the ground-state eigenvalue is excluded.

Finally, the neutral and charge-\(3e\) correlation lengths
increase near \(U_c\simeq12.0\)--\(12.5\)
[Fig.~\ref{fig:phase_transition}(b)].
Together with the changing charge response, this
identifies the candidate transition region. The
correlation-length growth motivates a continuous-transition
interpretation, whose limitations are assessed through
bond-dimension and circumference dependence.

\subsection{Entanglement spectrum}

To characterize the interaction-driven phase, we examine
the entanglement spectrum at a real-space bipartition
of the infinite cylinder. For Schmidt coefficients
\(\lambda_\alpha\), the entanglement energies are
\[
\epsilon_\alpha=-\ln\lambda_\alpha^2.
\]
The Schmidt states can be classified by their
\(\mathrm{SU}(3)\) representations.

An \(\mathrm{SU}(3)_1\) CSL has three topological sectors,
associated with
\(\mathbf{1}\), \(\mathbf{3}\), and \(\bar{\mathbf{3}}\).
Its chiral edge is described by the
\(\mathrm{SU}(3)_1\) Wess--Zumino--Witten (WZW)
conformal field theory. The predicted multiplets in the vacuum sector, organized by successive conformal levels, are distributed across momentum sectors, are
\begin{equation}
\begin{split}
&\mathbf{1},\\
&\mathbf{8},\\
&\mathbf{1}\oplus2\,\mathbf{8},\\
&2\,\mathbf{1}\oplus2\,\mathbf{8}
 \oplus\mathbf{10}\oplus\bar{\mathbf{10}},\ldots .
\end{split}
\end{equation}
Through these levels, the cumulative multiplicities are
four singlets, six octets, one \(\mathbf{10}\),
and one \(\bar{\mathbf{10}}\).

\begin{figure}[!t]
    \centering
    \includegraphics[width=0.9\columnwidth]{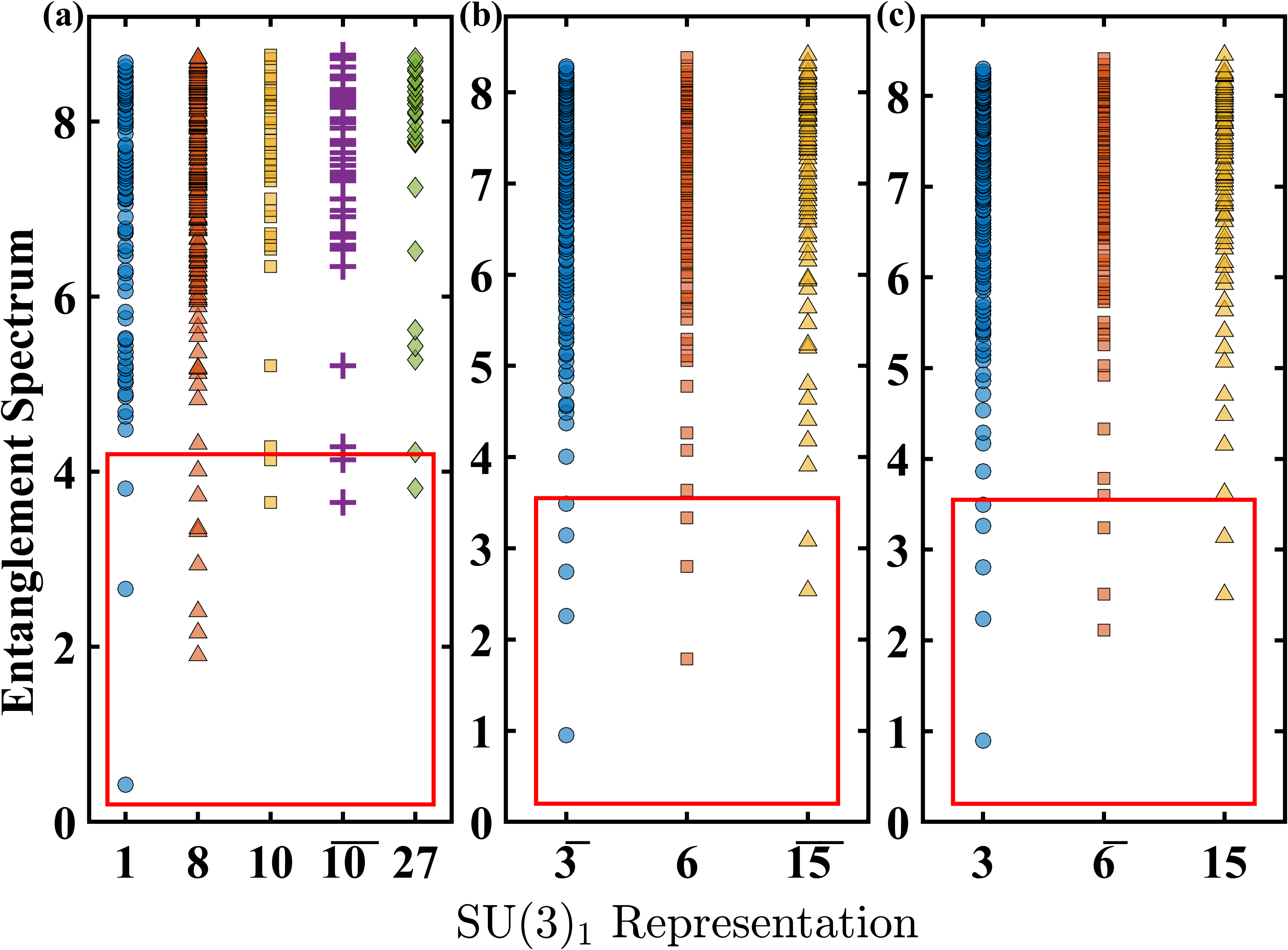}
    \caption{
    Representation-resolved entanglement spectra in Real space of \(\mathrm{SU}(3)_1\) CSL phase:
    (a) vacuum, (b) \(\bar{\mathbf{3}}\), and
    (c) \(\mathbf{3}\) topological sectors.
    The low-lying levels are labeled by
    \(\mathrm{SU}(3)\) representations.
    Their multiplet organization is compared with the
    corresponding \(\mathrm{SU}(3)_1\) WZW conformal
    towers. The red box is for convenience of view only.}
    
    \label{fig:entanglement_spectrum}
    \label{fig:ee_csl}
\end{figure}

Figure~\ref{fig:ee_csl} shows the numerical
representation-resolved spectra. The multiplet
organization is consistent with the expected CSL
edge structure. Momentum is not resolved in the
spectra presented here, so individual conformal
levels cannot be assigned solely from their
entanglement-energy ordering. Such ordering can
also be affected by overlap between neighboring
levels, as discussed in
Ref.~\cite{Arildsen2024Entanglement}.

For the \(\mathbf{3}\) sector, the first three
conformal levels contain
\[
\mathbf{3},\qquad
\mathbf{3}\oplus\bar{\mathbf{6}},\qquad
2\,\mathbf{3}\oplus\bar{\mathbf{6}}\oplus\mathbf{15}.
\]
The cumulative multiplicities are therefore four
\(\mathbf{3}\) multiplets, two \(\bar{\mathbf{6}}\)
multiplets, and one \(\mathbf{15}\).
The \(\bar{\mathbf{3}}\) sector has the conjugate
multiplet structure. Further details are given in
Appendix~\ref{subsec:SU3_1CFT}.

\subsection{Sector-resolved correlation lengths}

\begin{figure}[!t]
    \centering
    \includegraphics[width=0.9\columnwidth]
    {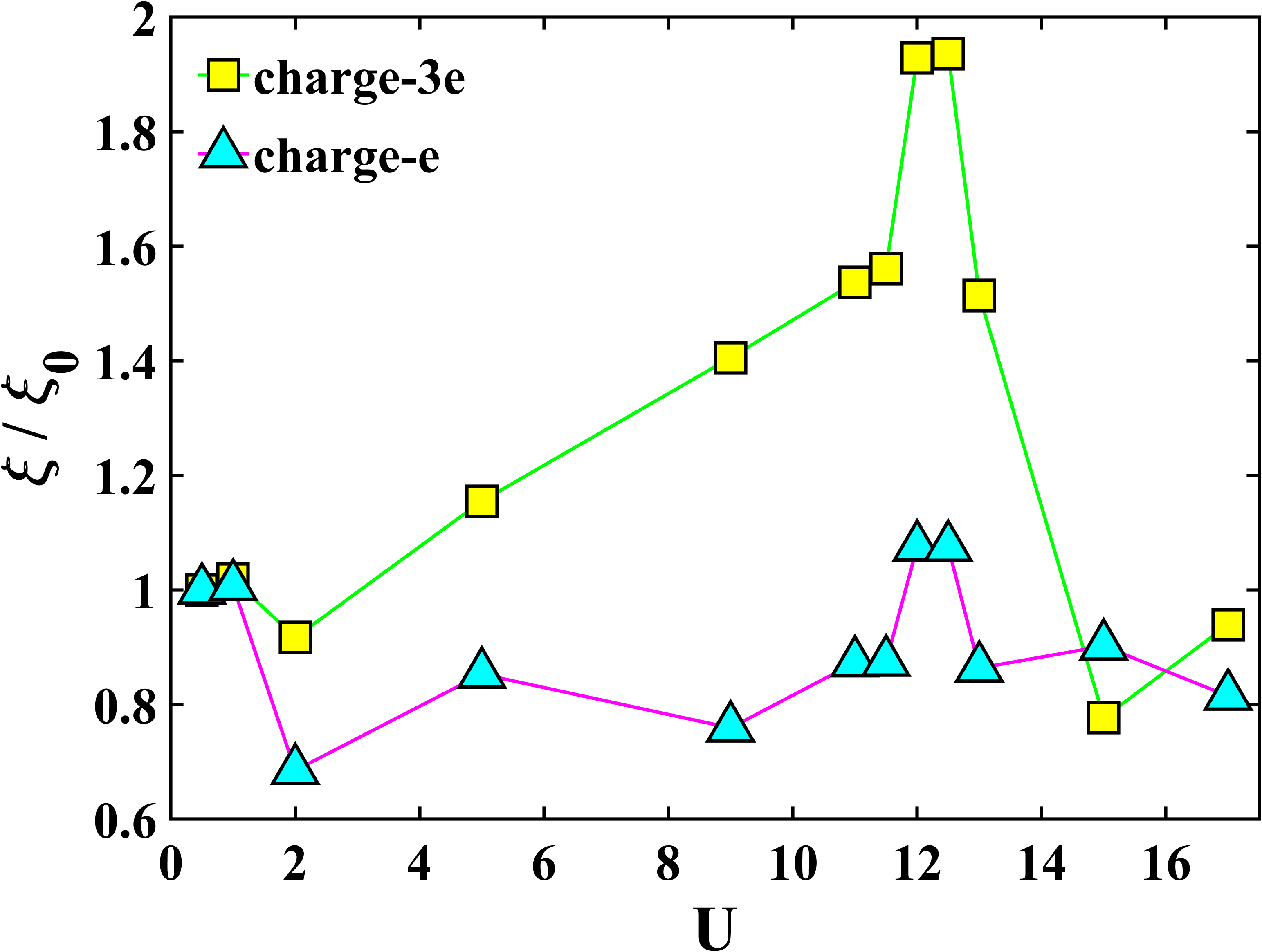}
    \caption{
    Interaction dependence of the charge-\(3e\) and
    charge-\(e\) correlation lengths.
    Squares denote the charge-\(3e\) sector and
    triangles the charge-\(e\) sector.
    Where normalized values are shown,
    \(\xi_0\) is the value at the smallest \(U\)
    in the corresponding dataset.
    The charge-\(3e\) correlation length increases
    near the transition, whereas the single-electron
    correlation length shows no comparable enhancement.
    }
    \label{fig:3e1e}
\end{figure}

We further confirm that as one tunes closer to the critical point, the correlation length of the \(3e\) sector increases, while the correlation length of the \(1e\) sector remains short-ranged. To analyze the data more clearly, we adopt the ratio of correlation lengths to examine the growth amplitude, namely \(\xi / \xi_0\), where \(\xi_0\) is the correlation length at the smallest \(U\) within the same dataset. Figure~\ref{fig:3e1e} compares the charge-\(e\) and
charge-\(3e\) sectors. To display their relative
enhancement, we normalize each correlation length
by its value \(\xi_0\) at the smallest \(U\) in the
corresponding dataset.

The single-electron correlation length remains short
throughout the transition region and decreases slightly
with increasing \(U\). This behavior is consistent
with a nonzero single-electron gap. Equal-time
correlation lengths alone, however, do not determine
the numerical value of an excitation gap.

The charge-\(3e\) correlation length behaves differently:
it increases markedly on approaching the transition
from the IQH side. The reported enhancement reaches
approximately \(200\%\) relative to the reference value.
This contrast shows that the trion channel develops
longer-ranged correlations without a corresponding
enhancement of single-electron correlations.

Within the candidate theory of
Eq.~\eqref{eq: critical}, this behavior is naturally
associated with the gauge-invariant charge-\(3e\)
monopole. 
Establishing an actual divergence
requires control of both finite bond dimension
and cylinder circumference.

\subsection{Interaction and bond-dimension dependence}

\begin{figure*}[htp]
    \centering
    \includegraphics[width=0.98\textwidth]
    {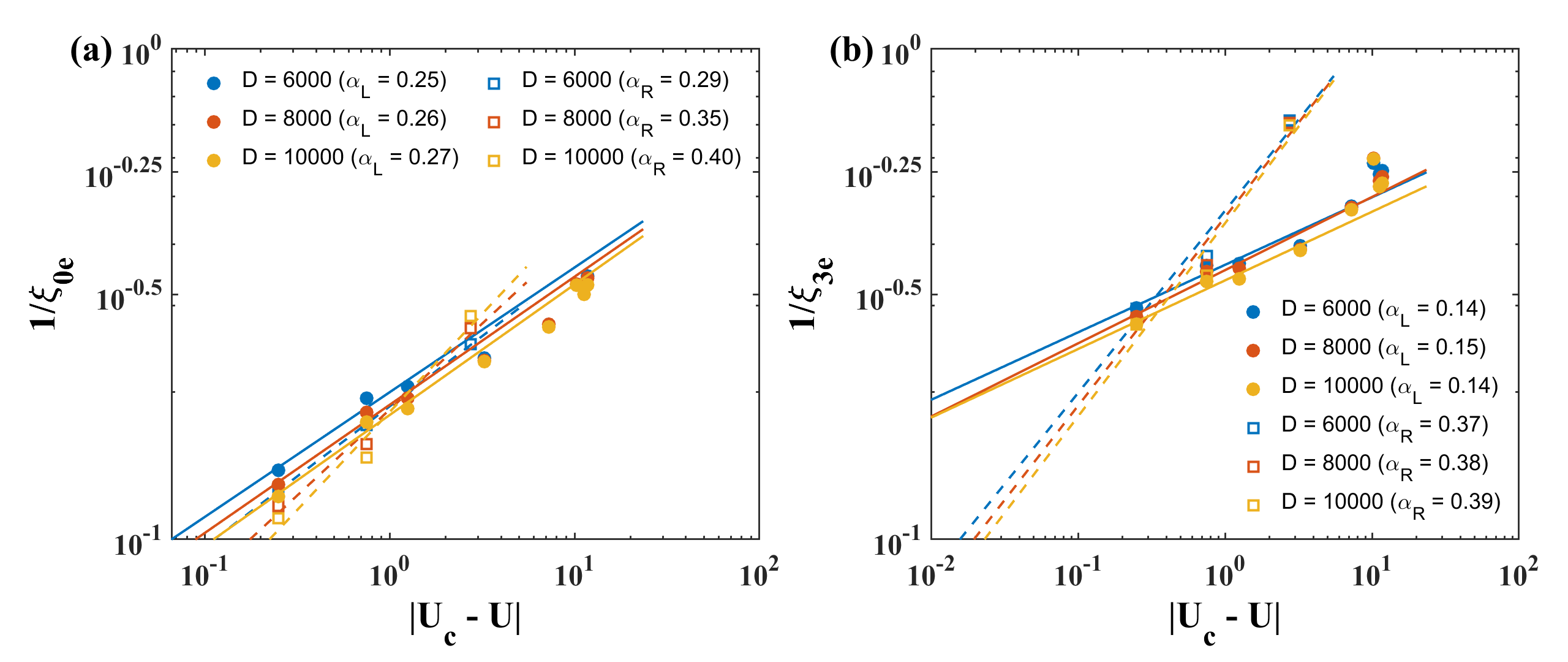}

    \medskip

    \includegraphics[width=0.98\textwidth]
    {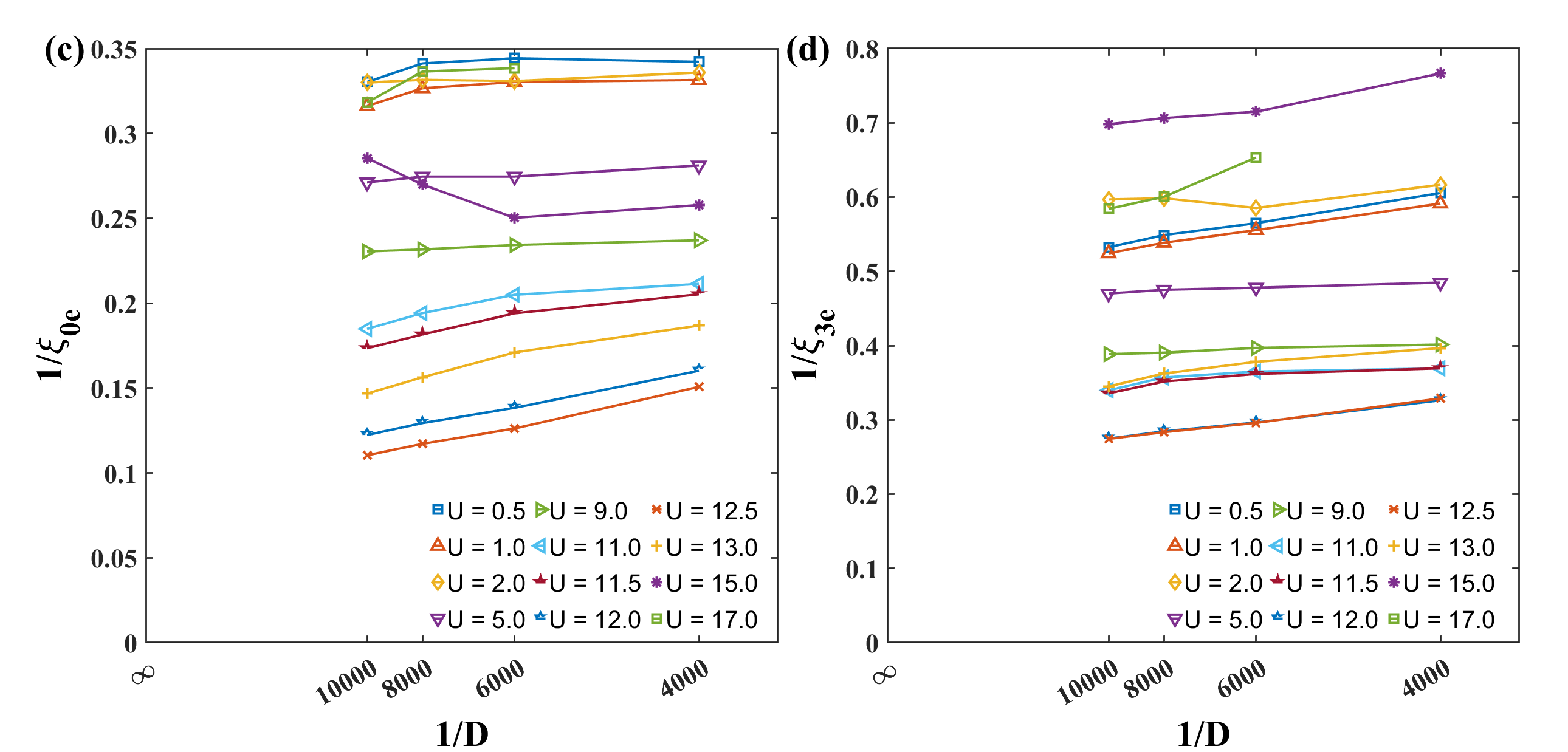}
    \caption{
    Correlation-length analysis near the candidate
    IQH--CSL transition.
    Top: interaction dependence of the inverse
    correlation lengths in (a) the neutral and
    (b) charge-\(3e\) sectors.
    Bottom: bond-dimension dependence for the
    corresponding sectors.
    The fits quantify the growth of correlation
    lengths and assess finite-entanglement effects.
    Extrapolation in bond dimension is performed
    at fixed cylinder circumference.
    }
    \label{fig:invcorr_vs_U}
    \label{fig:corr_vs_invD}
    \label{fig:scaling}
    \label{fig:group2}
\end{figure*}

To characterize the growth of the neutral and
charge-\(3e\) correlation lengths, we fit their inverse
values on the two sides of the transition according to,
\begin{equation}
\label{eq:interaction_scaling}
\xi_q^{-1}
=C_{q,\pm}|U-U_c|^{\alpha_{q,\pm}},
\end{equation}
where \(q\) labels the sector and \(\pm\) distinguish
the two sides,  as shown in the upper panel of Fig.~\ref{fig:group2}. The logarithmic form is
\[
\ln\xi_q^{-1}
=\ln C_{q,\pm}
+\alpha_{q,\pm}\ln|U-U_c|.
\]
We treat the fitted powers as effective exponents
over the accessible numerical range rather than
established universal critical exponents.

The fits at \(D=6000\), \(8000\), and \(10000\)
give the values summarized in
Table~\ref{tab:effective_exponents}.
On the IQH side, the neutral-sector exponent changes
from \(0.25\) to \(0.27\), while the charge-\(3e\)
exponent remains near \(0.14\)--\(0.15\).
On the CSL side, the neutral-sector exponent has
a stronger bond-dimension dependence, increasing
from \(0.29\) to \(0.40\); the charge-\(3e\) values
range from \(0.37\) to \(0.39\).

\begin{table}[t]
\centering
\begin{tabular}{c|ccc}
\hline\hline
Sector and side & \(D=6000\) & \(D=8000\) & \(D=10000\)\\
\hline
Neutral, IQH & 0.25 & 0.26 & 0.27\\
Neutral, CSL & 0.29 & 0.35 & 0.40\\
Charge-\(3e\), IQH & 0.14 & 0.15 & 0.14\\
Charge-\(3e\), CSL & 0.37 & 0.38 & 0.39\\
\hline\hline
\end{tabular}
\caption{
Effective exponents from fits to
Eq.~\eqref{eq:interaction_scaling}.
These values describe the accessible finite-cylinder,
finite-bond-dimension data.
}
\label{tab:effective_exponents}
\end{table}

The differences between sectors and between the two
sides indicate that a common asymptotic scaling regime
has not yet been established. Possible contributions
include finite-entanglement effects, finite
circumference, corrections to scaling, and uncertainty
in \(U_c\). The fits nevertheless quantify the
distinct growth of the neutral and charged
correlation lengths.

We further examine the bond-dimension dependence as shown in the bottom of Fig.~\ref{fig:group2}.
Finite bond dimension can substantially limit
correlation lengths, making extrapolation important
near criticality\cite{Rams2018,Vanhecke2019}.
As an empirical fitting form, we use
\begin{equation}
\label{eq:bond_scaling}
\xi_{q,D}^{-1}
=A_q D^{-b_q}+\xi_{q,\infty}^{-1},
\end{equation}
where \(A_q\), \(b_q\), and \(\xi_{q,\infty}\)
are fit parameters. The limit \(D\to\infty\)
is taken at fixed cylinder circumference and
therefore does not itself establish the
two-dimensional limit.

Near the candidate transition, the data are
consistent with a decreasing extrapolated inverse
correlation length in both sectors. If
\(\xi_{q,\infty}^{-1}=0\), then
\[
\ln\xi_{q,D}^{-1}
=\ln A_q-b_q\ln D.
\]
Thus a linear log--log dependence is consistent
with a diverging correlation length as one moves close to the putative transition around $U_c/t=12.5$ in Fig.~\ref{fig:group2}(c,d), but a
vanishing logarithmic intercept is not required.
Over a finite range of \(D\), a small nonzero
\(\xi_{q,\infty}^{-1}\) can also be difficult
to distinguish from zero.

Taken together, the interaction and bond-dimension
dependence support a continuous-transition
interpretation involving neutral and charge-\(3e\)
fluctuations. Wider cylinders and larger bond
dimensions are needed to establish the asymptotic
critical behavior and constrain its universality
class.

\section{Discussion}
\label{sec:discuss}

We have investigated the interaction-driven transition
between an IQH insulator and a candidate
\(\mathrm{SU}(3)_1\) CSL in the triangular-lattice
Hofstadter--Hubbard model. Charge pumping identifies
the weak-coupling IQH response, while the
representation-resolved entanglement spectra support
the CSL interpretation at stronger coupling.
The neutral and charge-\(3e\) correlation lengths
increase near \(U\simeq12.0\)--\(12.5\), whereas
single-electron correlations remain short-ranged.

The main result is the separation between the
single-electron and composite charged sectors.
It supports a critical description in which
fractionalized charge degrees of freedom reorganize
while the flavor-carrying partons remain gapped.
In the candidate level-three Chern--Simons--Higgs
theory, a gauge-invariant monopole carries charge
\(3e\) and fermionic statistics, providing a natural
interpretation of the enhanced trion correlations.
This extends the sector-resolved perspective of
the \(\mathrm{SU}(2)\) study to an odd-charge,
fermionic composite channel. 


These findings also motivate the study of finite doping cases.
If singlet trions remain energetically favorable
charge carriers away from the transition, doping
could produce a compressible trion state.
Pairing of fermionic charge-\(3e\) trions would
then provide a possible route to charge-\(6e\)
superconductivity. Neither the stability of a
trion liquid nor its pairing follows from the
undoped correlation lengths alone; both require
a separate finite-density calculation. Other exotic possibilities also remain to be explored\cite{Zhang2025Holon}.

We also note that the topological order of the SU(3)\(_1\) CSL is equivalent to that of the \(\nu = -1/3\) fermionic Laughlin state\cite{Zhang2025Holon}. This establishes a duality between doping the SU(3)\(_1\) CSL and doping the \(\nu = -2/3\) FCI: This means that doping the CSL with holes will produce a state dual to doping the FCI with electrons~\cite{Shi2025Doping}. In the context of doping the \(\nu = -2/3\) FCI in twisted MoTe\(_2\), the CDW metal in our theory has \(\sigma_{xy} \approx 0\), while doping FCI yields a state with \(\sigma_{xy} \approx e^2/h\) due to a background integer quantum Hall effect. This may have implications for the interpretation of the recently reported signatures of superconductivity from doping the \(\nu = -2/3\) FCI in twisted MoTe\(_2\)~\cite{Xu2025}.

\bibliography{trion_su3}

\clearpage
\appendix

\section{Lattice geometry and iDMRG setup}
\label{sec:lattice}

We describe the triangular lattice using
\[
\mathbf{a}_1=a\hat{x},\qquad
\mathbf{a}_2=a\left(
-\frac{1}{2}\hat{x}+\frac{\sqrt{3}}{2}\hat{y}
\right).
\]
The corresponding reciprocal vectors satisfying
\(\mathbf{a}_i\cdot\mathbf{b}_j=2\pi\delta_{ij}\)
are
\[
\mathbf{b}_1=\frac{2\pi}{a}
\left(\hat{x}+\frac{1}{\sqrt{3}}\hat{y}\right),
\qquad
\mathbf{b}_2=\frac{4\pi}{\sqrt{3}a}\hat{y}.
\]
Lattice sites are located at
\[
\mathbf{r}=n_1\mathbf{a}_1+n_2\mathbf{a}_2.
\]

With this choice of primitive vectors, an AC cylinder
of circumference \(L_y a\sqrt{3}\) is obtained by
identifying
\[
\mathbf{r}\sim\mathbf{r}+\mathbf{C},
\qquad
\mathbf{C}=L_y(\mathbf{a}_1+2\mathbf{a}_2).
\]

%

The hopping phases are chosen to generate flux
\(\pi/3\) per triangle. In a general gauge, a magnetic
translation acts as
\begin{equation}
\label{eq:bc}
T_\mu c^\dagger_{\mathbf{r}}T_\mu^{-1}
=e^{i\lambda_\mu(\mathbf{r})}
c^\dagger_{\mathbf{r}+\mathbf{a}_\mu},
\end{equation}
where the position-dependent gauge phases
\(\lambda_\mu(\mathbf{r})\) are determined by the
chosen Peierls convention.

Threading a dimensionless charge flux \(\phi\)
through the cylinder can be implemented by
\[
c^\dagger_{\mathbf{r}+\mathbf{C}}
=e^{i\phi}c^\dagger_{\mathbf{r}}.
\]
The magnetic translation phases and the boundary
twist have distinct roles and should not be
identified with one another.


\section{Local \(\mathrm{SU}(3)\) structure and
entanglement-spectrum counting}
\label{sec:su3_formalism}

\subsection{Generators and occupation-resolved chirality}

For the full electronic Hilbert space, define
\[
S_{i;\alpha\beta}
=c^\dagger_{i,\alpha}c_{i,\beta},
\qquad \alpha,\beta=1,2,3.
\]
The flavor generators are
\begin{equation}
T_i^a
=\frac{1}{2}
\sum_{\alpha,\beta=1}^{3}
(\lambda^a)_{\alpha\beta}S_{i;\alpha\beta},
\qquad a=1,\ldots,8,
\end{equation}
where \(\lambda^a\) are the Gell-Mann matrices.
They satisfy
\[
[T_i^a,T_i^b]=if^{abc}T_i^c.
\]
In the singly occupied subspace,
\(S_{i;\alpha\beta}=|\alpha\rangle_i\langle\beta|_i\).

The local eight-dimensional Hilbert space separates
into four particle-number sectors,
\[
(n,\mathcal{R})=
(0,\mathbf{1}),\quad
(1,\mathbf{3}),\quad
(2,\bar{\mathbf{3}}),\quad
(3,\mathbf{1}).
\]
These are four occupation sectors, not four
individual basis states.

The three-site operator is
\begin{equation}
P_{123}
=\sum_{\alpha,\beta,\gamma=1}^{3}
S_{1;\alpha\beta}\otimes
S_{2;\beta\gamma}\otimes
S_{3;\gamma\alpha},
\end{equation}
and the chirality is
\[
\chi_{123}=i(P_{123}-P_{123}^{\dagger}).
\]
Restricted to one particle per site,
\(P_{123}\) is a cyclic flavor permutation.
Its orientation fixes the overall sign convention
for \(\chi_{123}\).

Let \(\Pi_i^{(n)}\) project onto local particle
number \(n\). The occupation-resolved chirality
contribution is defined by
\begin{equation}
\chi_{n_1n_2n_3}
=
\left\langle
\Pi_1^{(n_1)}\Pi_2^{(n_2)}\Pi_3^{(n_3)}
\chi_{123}
\Pi_1^{(n_1)}\Pi_2^{(n_2)}\Pi_3^{(n_3)}
\right\rangle.
\end{equation}
The sectors emphasized in the main text correspond
to \(n_1n_2n_3=111\) and to configurations with
two singly occupied sites and one doubly occupied
site.


\subsection{\(\mathrm{SU}(3)_1\) conformal towers}
\label{subsec:SU3_1CFT}

The low-lying \(\mathrm{SU}(3)_1\) WZW states can be
organized by multi-spinon mode sequences
\(\{n_i\}\) and global \(\mathrm{SU}(3)\)
multiplets\citeapp{Bouwknegt1996SU,Arildsen2024,
Chen2020SU3}. Table~\ref{tab:SU3_full} lists the
states through conformal weight \(L_0=3\).
We write repeated representations with explicit
multiplicities.

\begin{table}[h]
\centering
\small
\begin{tabular}{|c|c|c|}
\hline
\(L_0\) & \(\{n_i\}\) & \(\mathrm{SU}(3)\) multiplet \\
\hline
0 & --- & \(\mathbf{1}\) \\
\(1/3\) & 0 & \(\bar{\mathbf{3}}\) \\
\(1/3\) & 0,0 & \(\mathbf{3}\) \\
1 & 0,0,1 & \(\mathbf{8}\) \\
\(4/3\) & 1 & \(\bar{\mathbf{3}}\) \\
\(4/3\) & 0,0,1,1 & \(\mathbf{6}\) \\
\(4/3\) & 0,1 & \(\mathbf{3} \oplus \bar{\mathbf{6}}\) \\
2 & 0,1,1 & \(\mathbf{8}\) \\
2 & 0,0,2 & \(\mathbf{1} \oplus \mathbf{8}\) \\
\(7/3\) & 2 & \(\bar{\mathbf{3}}\) \\
\(7/3\) & 0,0,1,2 & \(\bar{\mathbf{3}} \oplus \mathbf{6} \oplus \mathbf{15}\) \\
\(7/3\) & 1,1 & \(\mathbf{3}\) \\
\(7/3\) & 0,2 & \(\mathbf{3} \oplus \bar{\mathbf{6}}\) \\
\(7/3\) & 0,0,1,1,2 & \(\mathbf{15}\) \\
3 & 0,0,3 & \(\mathbf{1} \oplus \mathbf{8}\) \\
3 & 0,1,2 & \(\mathbf{1} \oplus \mathbf{8}_2 \oplus \overline{\mathbf{10}}\) \\
3 & 0,0,1,1,2,2 & \(\mathbf{10}\) \\
\hline
\end{tabular}
\caption{\(\mathrm{SU}(3)_1\) WZW model: states for \(L_0 \le 3\).}
\label{tab:SU3_full}
\end{table}

The vacuum sector is obtained from sequences
containing a multiple of three spinons.
Its successive conformal levels contain
\(1\), \(1\), \(3\), and \(7\) global
\(\mathrm{SU}(3)\) multiplets.
Including the dimensions of those multiplets,
the corresponding state counts are
\(1\), \(8\), \(17\), and \(46\).
Multiplet counts must therefore be distinguished
from total state degeneracies.

\begin{table}[H]
\centering
\small
\begin{tabular}{|c|c|c|}
\hline
\(L_0\) & \(\{n_i\}\) & \(\mathrm{SU}(3)\) multiplet \\
\hline
0 & --- & \(\mathbf{1}\) \\
1 & 0,0,1 & \(\mathbf{8}\) \\
2 & 0,1,1 & \(\mathbf{8}\) \\
2 & 0,0,2 & \(\mathbf{1} \oplus \mathbf{8}\) \\
3 & 0,0,3 & \(\mathbf{1} \oplus \mathbf{8}\) \\
3 & 0,1,2 & \(\mathbf{1} \oplus \mathbf{8}_2 \oplus \overline{\mathbf{10}}\) \\
3 & 0,0,1,1,2,2 & \(\mathbf{10}\) \\
\hline
\end{tabular}
\caption{Vacuum conformal tower of the
\(\mathrm{SU}(3)_1\) theory.}
\label{tab:SU3_vacuum}
\end{table}

With the convention that an elementary spinon
transforms as \(\bar{\mathbf{3}}\), the
\(\mathbf{3}\) sector contains sequences with
spinon number \(N\equiv2\pmod3\).
The first three levels contain \(1\), \(2\),
and \(4\) multiplets, corresponding to total
state counts \(3\), \(9\), and \(27\).

\begin{table}[H]
\centering
\small
\begin{tabular}{|c|c|c|}
\hline
\(L_0\) & \(\{n_i\}\) & \(\mathrm{SU}(3)\) multiplet \\
\hline
\(1/3\) & 0,0 & \(\mathbf{3}\) \\
\(4/3\) & 0,1 & \(\mathbf{3} \oplus \bar{\mathbf{6}}\) \\
\(7/3\) & 1,1 & \(\mathbf{3}\) \\
\(7/3\) & 0,2 & \(\mathbf{3} \oplus \bar{\mathbf{6}}\) \\
\(7/3\) & 0,0,1,1,2 & \(\mathbf{15}\) \\
\hline
\end{tabular}
\caption{The \(\mathbf{3}\) conformal tower.}
\label{tab:SU3_3}
\end{table}

The conjugate \(\bar{\mathbf{3}}\) sector
contains sequences with \(N\equiv1\pmod3\)
in the same convention.
It has the conjugate multiplet structure
and identical total degeneracies.

\begin{table}[H]
\centering
\small
\begin{tabular}{|c|c|c|}
\hline
\(L_0\) & \(\{n_i\}\) & \(\mathrm{SU}(3)\) multiplet \\
\hline
\(1/3\) & 0 & \(\bar{\mathbf{3}}\) \\
\(4/3\) & 1 & \(\bar{\mathbf{3}}\) \\
\(4/3\) & 0,0,1,1 & \(\mathbf{6}\) \\
\(7/3\) & 2 & \(\bar{\mathbf{3}}\) \\
\(7/3\) & 0,0,1,2 & \(\bar{\mathbf{3}} \oplus \mathbf{6} \oplus \overline{\mathbf{15}}\) \\
\hline
\end{tabular}
\caption{The \(\bar{\mathbf{3}}\) conformal tower.}
\label{tab:SU3_3bar}
\end{table}

\section{Circumference dependence and charge pumping}
\label{sec:pump_charge_correlation_different_Ly}

\subsection{Additional correlation-length data}

\begin{figure}[htbp]
    \centering
    \includegraphics[width=0.98\columnwidth]
    {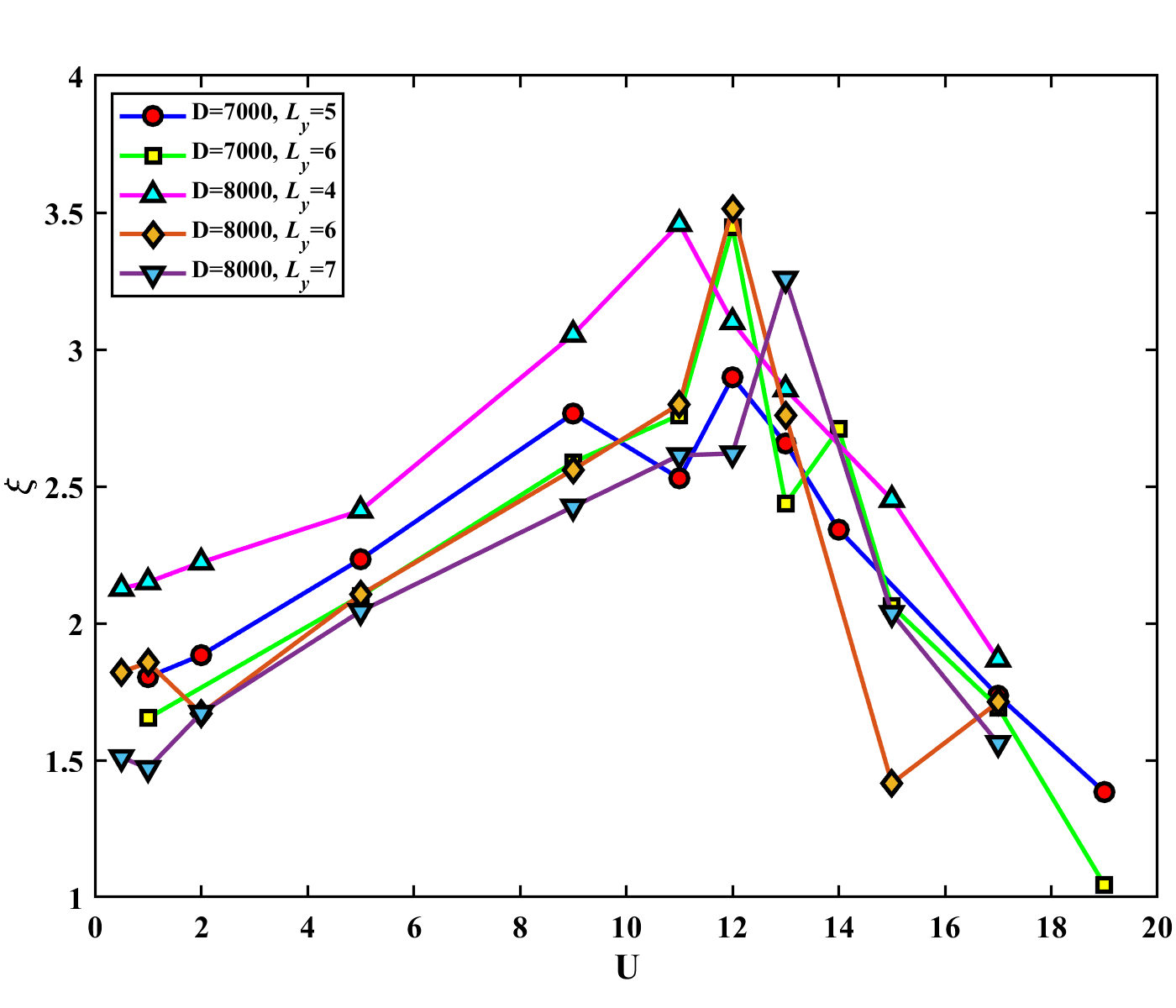}
    \caption{
    Supplementary charge-\(3e\) correlation-length  versus the interaction strength $U$
     for different cylinder circumferences
    and bond dimensions.
    The datasets include \(L_y=4,6,7\) at \(D=8000\)
    and \(L_y=5,6\) at \(D=7000\).
    }
    \label{fig:append1_corr_vs_L}
\end{figure}

Figure~\ref{fig:append1_corr_vs_L} compares
charge-\(3e\) correlation-length data obtained
for several cylinder circumferences and bond
dimensions. The datasets include \(L_y=4,6,7\)
at \(D=8000\), and \(L_y=5,6\) at \(D=7000\).
The enhancement near the transition is observed
in each dataset, indicating that it is not
restricted to a single circumference or bond
dimension. These comparisons do not, by
themselves, constitute a controlled
two-dimensional extrapolation.


\subsection{Charge pumping}

We probe the charge response by adiabatically
threading flux \(\Psi\) through the cylinder.
At a bipartition into two semi-infinite
cylinders, let \(\lambda_\alpha(\Psi)\)
be the Schmidt coefficients and \(Q_\alpha\)
their charge labels. The change in the
Schmidt-weighted charge is
\begin{equation}
\label{eq:pumped_charge}
\begin{split}
\Delta Q(\Psi)
={}&
\sum_\alpha Q_\alpha
|\lambda_\alpha(\Psi)|^2
\\
&-
\sum_\alpha Q_\alpha
|\lambda_\alpha(0)|^2,
\end{split}
\end{equation}
with the charge-label convention tracked
consistently during flux insertion.

For a dimensionless flux \(\Psi\) of period
\(2\pi\), insertion of one flux quantum pumps
charge
\[
\Delta Q(2\pi)
=\frac{\sigma_{xy}}{e^2/h}\,e,
\]
up to the sign set by the orientation convention.
Thus the weak-coupling IQH phase pumps charge
of magnitude \(3e\) per flux quantum.

\begin{figure}[htbp]
    \centering
    \includegraphics[width=0.98\columnwidth]
    {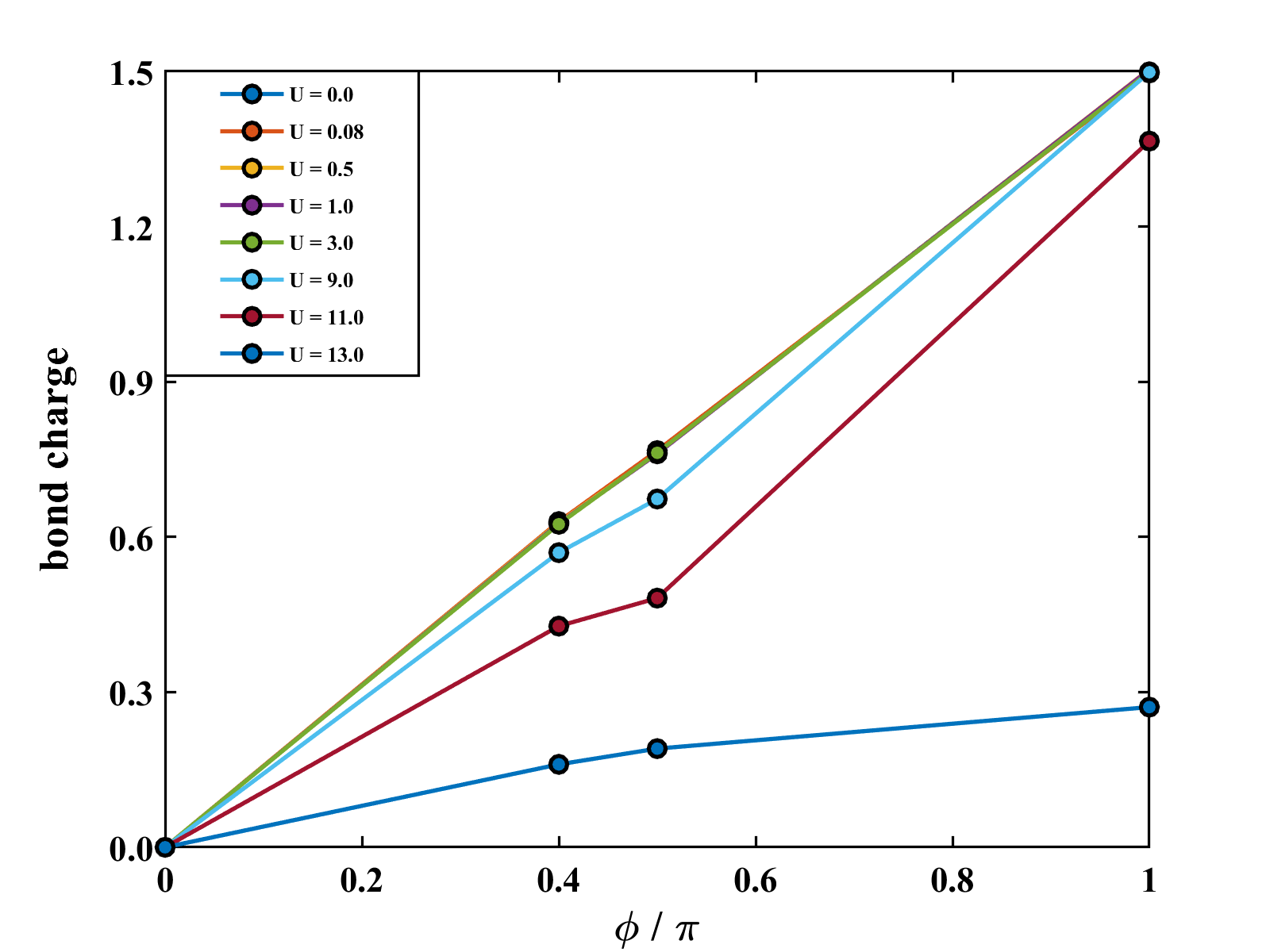}
    \caption{
    Pumped charge \(\Delta Q(\Psi)\) under
    flux insertion for different \(U\).
    At weak coupling, insertion of one flux
    quantum pumps charge of magnitude \(3e\),
    consistent with \(\sigma_{xy}=3e^2/h\).
    The response changes near the candidate
    IQH--CSL transition.
    }
    \label{fig:append2_pump_charge}
\end{figure}

Figure~\ref{fig:append2_pump_charge} shows the
pumping curves at different interaction
strengths. Their departure from the
weak-coupling response is used together
with the other diagnostics to locate the
transition region.


\newpage
\section*{Appendix References}

\end{document}